\documentclass[10pt]{article}
\usepackage{graphicx}
\usepackage{amsmath}

\def\Title#1{\begin{center} {\Large #1 } \end{center}}
\def\Author#1{\begin{center}{ \sc #1} \end{center}}
\def\Address#1{\begin{center}{ \it #1} \end{center}}

\newenvironment{Abstract}{\begin{quotation} \begin{center} 
             \large ABSTRACT \end{center}\bigskip 
      \begin{center}\begin{large}}{\end{large}\end{center} \end{quotation}}

\newenvironment{Presented}{\begin{quotation} \begin{center} 
             PRESENTED AT\end{center}\bigskip 
      \begin{center}\begin{large}}{\end{large}\end{center} \end{quotation}}

\def\Acknowledgements{\bigskip  \bigskip \begin{center} \begin{large}
             \bf ACKNOWLEDGEMENTS \end{large}\end{center}}
\def\beq{\begin{equation}}
\def\eeq#1{\label{#1}\end{equation}}
\def\eeqn{\end{equation}}

\def\beqa{\begin{eqnarray}}
\def\eeqa#1{\label{#1}\end{eqnarray}}
\def\eeqan{\end{eqnarray}}

\let\bar=\overbar

\def\Dslash{\not{\hbox{\kern-4pt $D$}}}
\def\dslash{\not{\hbox{\kern-2pt $\del$}}}

\def\msb{{\bar{\ssstyle M \kern -1pt S}}}

\usepackage{color}

\def\affiliation{
University of Helsinki and Helsinki Institute of Physics, \\
P.O. Box 64, FI-00014, Helsinki, Finland,\\ 
on behalf of the CMS, LHCf and TOTEM Collaborations}

\begin{document}

% large size for the first page
\large
\begin{titlepage}
%\pubblock

%% Change the title, name, abstract
%% Title 
\vfill
\Title{  Very forward measurements at the LHC  }
\vfill

%  if you need to add the support use this, fill the \support definition above. 
%   \Author{ FIRSTNAME LASTNAME \support }
\Author{ Mirko Berretti  }
\Address{\affiliation}
\vfill
\begin{Abstract}
In this talk we present a selection of forward physics results recently obtained with the run-1 and run-2 LHC data by the CMS, LHCf and TOTEM experiments.
The status of the very forward LHC proton spectrometer, CT-PPS, is discussed: emphasis is given to the physics potential of CT-PPS and to the analyses that are 
currently ongoing with the data collected in 2016.
Very recent forward measurements obtained with the LHCf and the CMS-CASTOR calorimeter are then addressed. In particular, CMS measured the inclusive energy spectrum in 
the very forward direction for proton-proton collisions at a center-of-mass energy of 13 TeV and the jet cross sections for p+Pb collisions at 5.02 TeV. The LHCf 
experiment has instead recently published the inclusive energy spectra of forward photons for pp collisions at 13 TeV.
Finally, the new measurements of the total, elastic and inelastic cross sections obtained by the TOTEM collaboration at 2.76 and 13 TeV center of mass energy are presented.
\end{Abstract}
\vfill

% DO NOT CHANGE 
\begin{Presented}
The Fifth Annual Conference\\
 on Large Hadron Collider Physics \\
Shanghai Jiao Tong University, Shanghai, China\\ 
May 15-20, 2017
\end{Presented}
\vfill
\end{titlepage}
\def\thefootnote{\fnsymbol{footnote}}
\setcounter{footnote}{0}
%

% normal size for the rest
\normalsize 

%% Your paper should be entered below. 
\section{Physics results and status of CT-PPS at the LHC}
The CMS-TOTEM Precision Proton Spectrometer (CT-PPS) \cite{ctpps1} is a joint CMS and TOTEM project that focuses on the study of rare central
exclusive (CE) reactions, which can occur at the LHC in photon-photon ($\gamma\gamma$),
Pomeron-Pomeron or $\gamma$-Pomeron collisions. 

\begin{figure}[htb]
\centering
\includegraphics[height=1.5in]{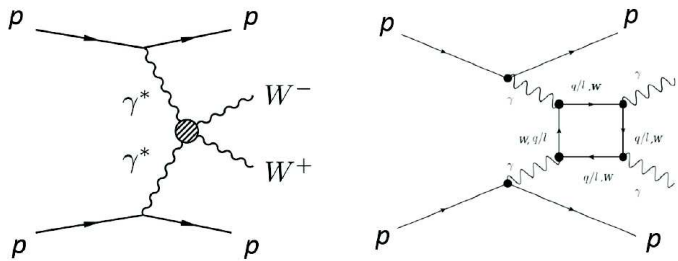}
\caption{Exclusive WW (left) and $\gamma\gamma$ (right) production.}
\label{fig1}
\end{figure}

CT-PPS has the potential to constrain the vector boson anomalous quartic gauge couplings and measure the exclusive production of 
$\gamma\gamma$, WW and ZZ (fig. \ref{fig1}) with better precision than that obtained by CMS or ATLAS without proton tagging. The spectrometer 
consists of two arms, one on each side of the interaction point (IP). Each arm has two tracking stations, one instrumented with TOTEM silicon strips 
and one with 3D pixels  \cite{ctpps2}; it also has one timing station equipped with 3 scCVD diamonds planes \cite{ctpps3} and one UFSD plane \cite{ctpps4}. 
The stations will measure protons produced in CE events with masses from ∼300 GeV up to ∼2 TeV. All stations are housed in near-beam Roman Pots.

\begin{figure}[htb]
\centering
\includegraphics[height=2.5in]{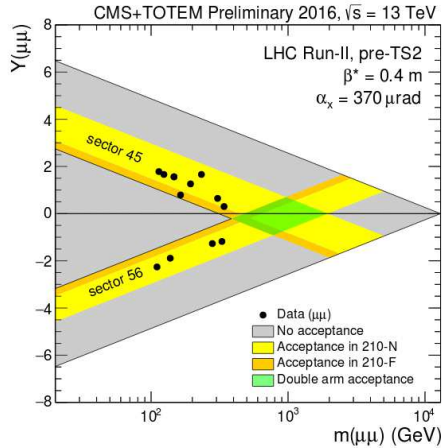}
\caption{Expected coverage in the rapidity vs invariant mass plane overlaid with the observed dimuon signal candidate events \cite{ctpps5}.}
\label{figctpps}
\end{figure}

Timing detectors are necessary to reconstruct the longitudinal position of the vertex from which the scattered protons originate, 
so that they can be unambiguously associated to one of the vertices reconstructed by the CMS central detectors even in high pileup runs. 
The spectrometer started to take data in June 2016 and collected about 15 fb$^{-1}$ in 2016. In 2017, it has taken already an additional 10 fb$^{-1}$ both with 
the tracking and the timing detectors. This data sample constitutes the largest CE sample ever collected in high energy physics. 
Several analyses are currently ongoing; among them, the search for exclusive dileptons or $\gamma\gamma$, as well as the measurement of CE events with large missing mass. 
The process  pp$\rightarrow$ p$\mu^{+}\mu^{-}$p$^{*}$ has been observed for dimuon masses larger than 110 GeV in pp collisions at $\sqrt{s}=13$ TeV \cite{ctpps5}.
Here p$^{*}$ indicates that the second proton is undetected, and either remains intact or dissociates into a low-mass state p$^{*}$. 
The other intact proton is instead measured in CT-PPS.  A total of 12 candidates with m($\mu\mu$)$>$110 GeV, and matching forward proton kinematics, is observed.
The value of the dimuon mass and rapidity is presented in fig. \ref{figctpps} together with the mass-rapidity detector coverage. 
This corresponds to an excess of more than four standard deviations over the background. This result constitutes the first evidence of this electromagnetic process 
at such masses and it also demonstrates that CT-PPS performs as expected.

\section{New CMS results on forward energy flow in pp collisions}
\begin{figure}[htb]
\centering
\includegraphics[height=3in]{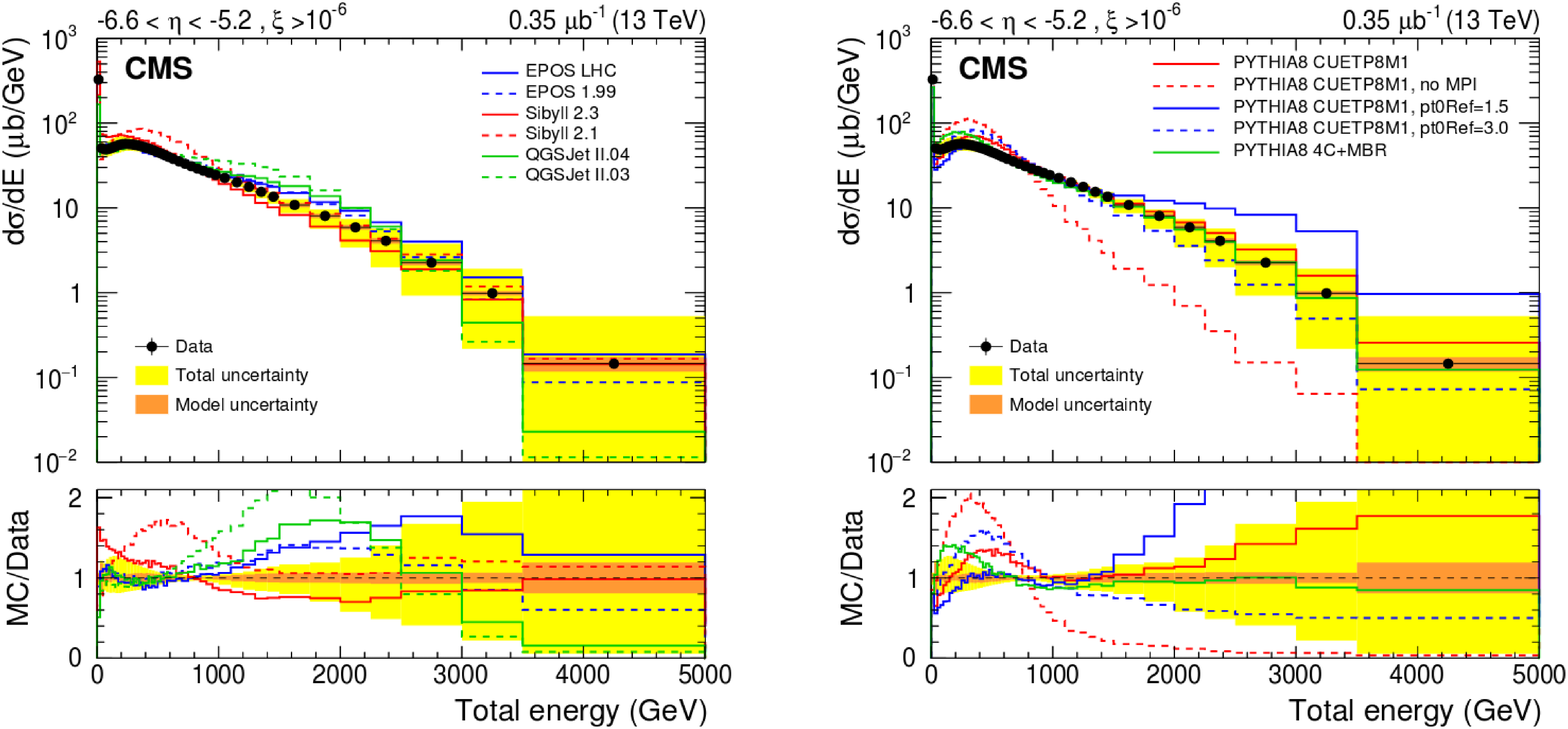}
\caption{Differential cross section as a function of the total energy in the region
-6.6 $<\eta<$ -5.2 \cite{cmsE1}. The left panel shows the data compared to MC event generators mostly developed for cosmic ray induced air showers, while in the right panel 
different PYTHIA 8 tunes are compared.}
\label{fig2}
\end{figure}

CMS \cite{cmsdet} has recently published \cite{cmsE1} the measurement of the differential cross section as function of the energy for inclusive particle production in the 
pseudorapidity ($\eta$) region -6.6 $<\eta<$ -5.2. The measurement is obtained with 0.35 $\mu b^{-1}$ of proton-proton collisions data at a center of mass energy of 13 TeV, 
using the CASTOR calorimeter. The cross section is given as a function of the total energy deposited in CASTOR (see fig. \ref{fig2}), as well as of its electromagnetic and hadronic 
components \cite{cmsE1}. 

The spectra are sensitive to  the modeling of multiparton interactions in pp collisions, and provide new constraints for hadronic interaction models used in 
collider and in high energy cosmic ray physics. PYTHIA8 CUETP8M1 without MPI is ruled out by the data, which exhibit much
harder spectra than predicted by the model.  Moreover, the shape of the spectra are significantly influenced by the MPI-related settings in PYTHIA8.  Event generators developed for modeling high energy cosmic ray air showers, tuned to LHC
measurements at 0.9, 7, and 8 TeV, agree better with the present data than those tuned to Tevatron results alone. This is especially true for QGSJETII and SIBYLL.
 
\section{New CMS results on forward jet production in pA collisions}
CMS has recently published \cite{cmsE2} the measurement of the very forward inclusive jet cross sections in p-Pb collisions at a center 
of mass energy of $\sqrt{s_{NN}}=$ 5.02 TeV. Jets are measured with the CASTOR calorimeter, in the pseudorapidity region -6.6 $<\eta<$ -5.2. 
p-Pb collisions are ideal to search for signatures of non-DGLAP parton evolution, as effects are expected to be important at small values of $x$\footnote{$x$ is the  
longitudinal momentum fraction carried by the parton.} and high parton densities. 
This regime can be sensitive to non-linear parton dynamics (i.e. gluon saturation) where the gluon density stops rising because of the high rate of recombination processes.
In ions, gluon densities are larger than in protons; moreover, very small values of $x$ can be accessed given the high rapidity of CASTOR.

\begin{figure}[htb]
\centering
\includegraphics[height=3.5in]{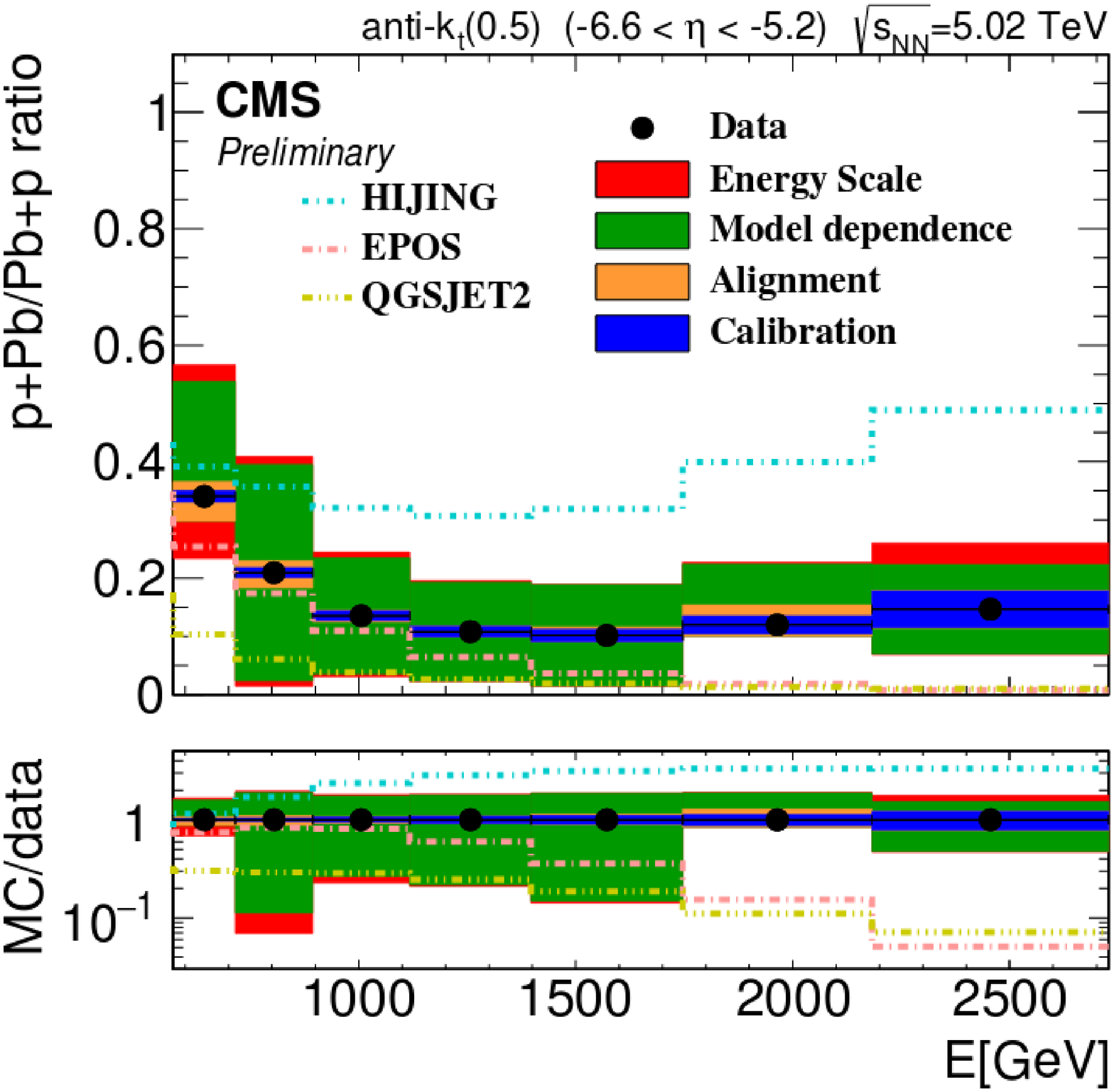}
\caption{The ratio of the differential CASTOR jet energy cross sections for p+Pb and Pb+p \cite{cmsE2}.  The
CASTOR  jets  were  unfolded  to particle  level. Model predictions  are included for EPOS-LHC, HIJING, and QGSJETII-04.}
\label{fig3}
\end{figure}

The data samples correspond to integrated luminosities of 3.13 and 6.71 $nb^{-1}$, for beam configurations with the incoming proton (p+Pb) or incoming ion (Pb+p) towards CASTOR, respectively.
The spectra are unfolded to the particle level and compared to predictions of various event generators.
In addition, the ratio of the spectra in p+Pb and Pb+p collisions is measured (see fig. \ref{fig3}); in the ratio, many systematic uncertainties cancel, notably that due the energy scale.
The results are corrected in order to be representative of events having a particle on both sides of the HF acceptance with a minimal energy of 4 GeV and at least a 
charged particle in the tracker acceptance with p$_T$ above 0.4 GeV/c.
None of the event generators under consideration describes all the measured spectra in their full energy range, and deviations
between data and models of more than two orders of magnitude are observed. Specifically, the Pb+p spectrum is underestimated at lower energy, 
while the models are consistent with the data for E $>$ 1.2 TeV. All the models fail to describe the p+Pb/Pb+p ratio. However, the p+Pb spectrum is well described by HIJING.

\section{Latest LHCf results}
The LHCf experiment has recently measured the forward photon-energy spectra in proton-proton collisions at $\sqrt{s} =$ 13 TeV \cite{lhcf1}.
High energy photons are measured on both sides of the interaction point in the region at $|\eta|>$ 10.94 and 8.81 $<|\eta|<$ 8.99. 

\begin{figure}[htb!]
\centering
\includegraphics[height=3.5in]{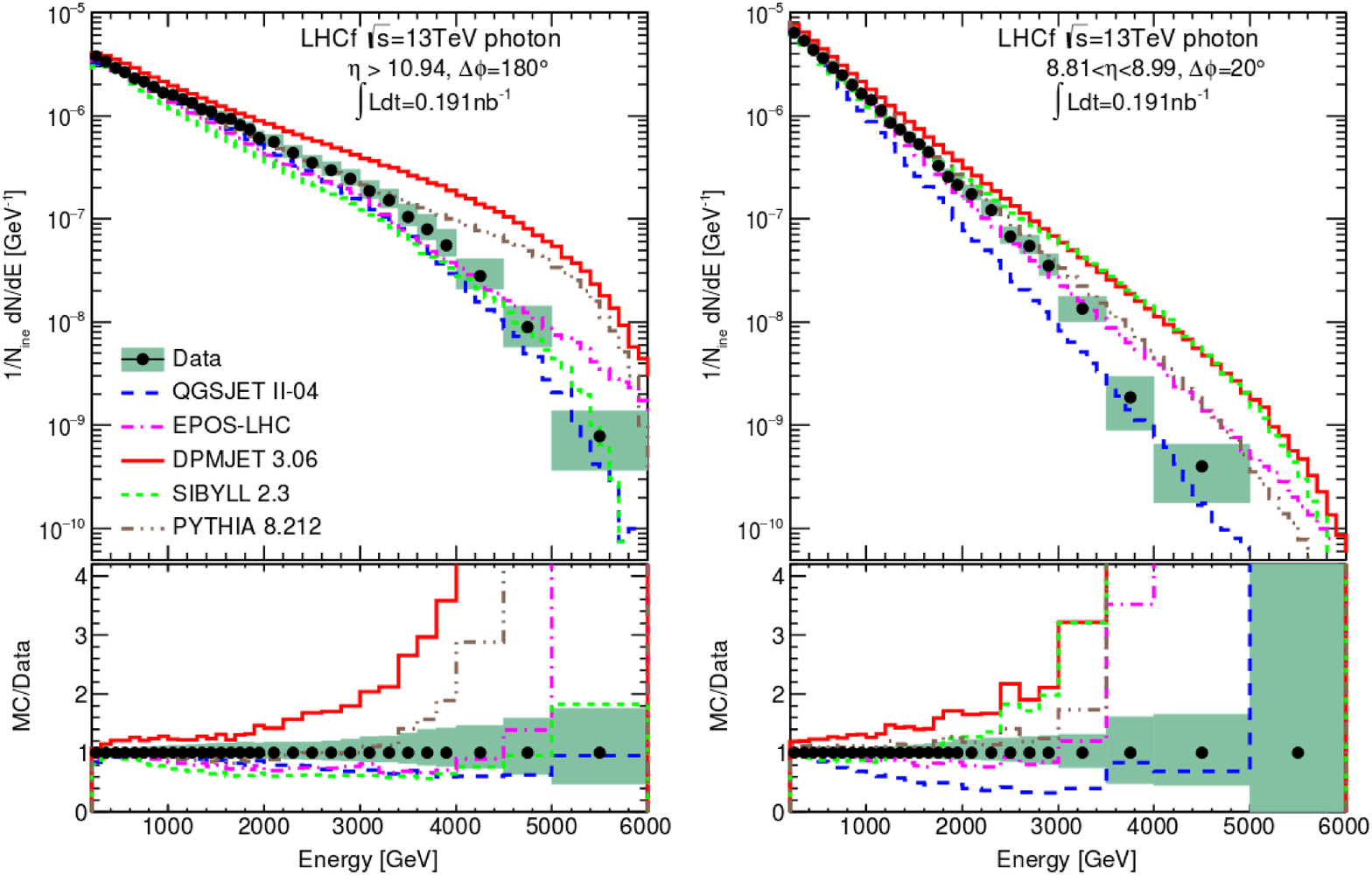}
\caption{Photon spectra obtained with the LHCf data at $\sqrt{s} =$ 13 TeV \cite{lhcf1} and MC predictions. 
The top panels show the energy spectra, the bottom panels the ratio of MC predictions to the data. The hatched areas indicate the 
total uncertainties of experimental data including the statistical and the systematic uncertainties.}
\label{flhcf1}
\end{figure}

This measurement is very sensitive to the very forward production of $\pi^{0}$ mesons, and is thus important to improve hadronic interaction models used 
in the analysis of high energy cosmic ray showers. The measured inclusive photon energy spectrum is shown in fig. \ref{flhcf1}, along with predictions 
from MC generators frequently used in cosmic ray physics. Although none of the models reproduces perfectly the data, EPOS-LHC exhibits the best agreement.

The differential cross section as a function of the energy has been measured also for neutron production, in pp collision at $\sqrt{s} =$ 13 TeV \cite{lhcf2}. 
The MC models are lower than the data for  $|\eta|>$ 10.76. LHCf and the ATLAS experiment, which share the same IP, operated jointly in 2015 \cite{lhcf3}. 
This allows detailed studies of the very forward production of high energy neutral particles in conjunction with event-by-event information on 
particle production in the central region.

\section{Latest TOTEM results}

\begin{figure}[htb!]
\centering
\includegraphics[height=3in]{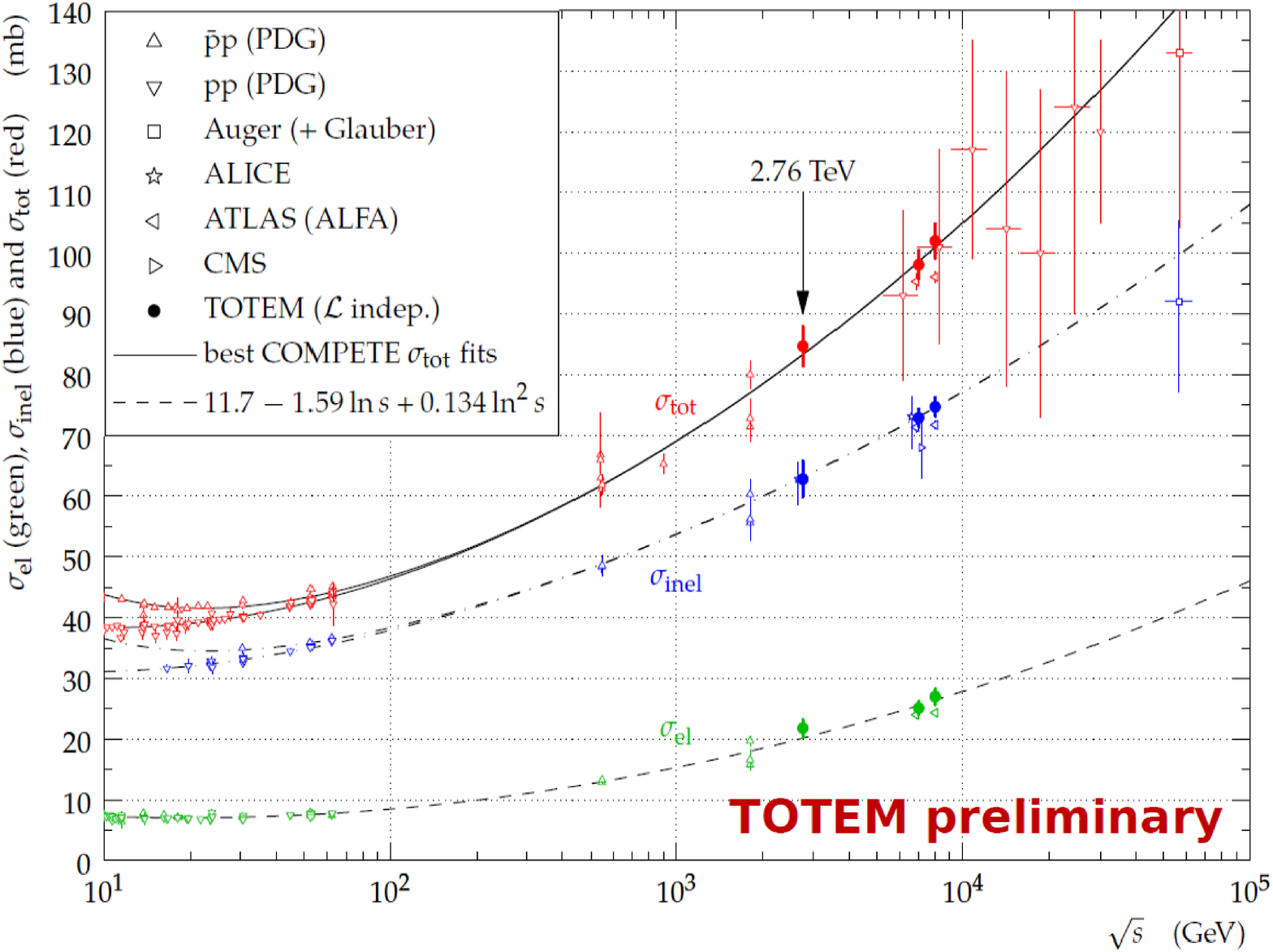}
\caption{Measurements of the total, elastic and inelastic pp cross sections as a function of the center of mass energy, including the new preliminary results at 2.76 TeV (adapted from \cite{totemAll}).}
\label{totemfig}
\end{figure}

The TOTEM experiment has recently performed detailed  
 studies of the $t$\footnote{$t$ is the square of the four-momentum transfer between the protons} distribution in elastic pp scattering at 13 TeV, 
including also the nuclear-Coulomb interference region at very low $|t|$. 
This measurement makes it possible to extract the  $\rho$ parameter ($\rho$ is the ratio between the real 
and imaginary part of the elastic amplitude at $t=$0) with unprecedented precision.
The study of the elastic slope parameter $B$ ($B=d(\text{ln}\, d\sigma/dt)/dt|_{t=0}$) 
as a function of the center of mass energy (2.76, 7, 8, 13 TeV) shows that $B$ grows faster than $\text{ln}s$, a behavior that can be due to multipomeron contributions \cite{totem0}.
The measurement of the total, inelastic and elastic cross sections has also been performed at the center of mass energy of 2.76 TeV. 
The TOTEM results favor the CDF measurement \cite{totem1} of the proton-antiproton total cross section at 1.8 TeV, but are not incompatible with 
the E811 results \cite{totem2}. The preliminary measurements of the elastic, inelastic and total pp cross sections at 2.76 TeV are included in fig. \ref{totemfig}.

\Acknowledgements
I am grateful to the organizers of LHCP2017 for their kind invitation and for the very stimulating and well organized conference.
I appreciated the help from the CMS and LHCf colleagues in the selection of the results shown at this conference.

\end{document}